\documentclass[global,twocolumn]{svjour}
\usepackage{graphicx}
\journalname{myjournal}
\usepackage{amsmath}
\usepackage{hyperref}

\begin{document}
  \title{Complete characterization of weak, ultrashort near-UV pulses by spectral interferometry.}
  \author{Marcin Kacprowicz\thanks{tel:~+48~56~6113332, fax:~+48~56~6225397, E-mail:~mkacpr@gmail.com}, Wojciech Wasilewski and Konrad Banaszek}
  \institute{Institute of Physics, Nicolaus Copernicus University, Grudzi\c{a}dzka 5, 87-100 Toru\'{n}, Poland}
 \date{\today}
  \maketitle
\begin{abstract}
  We present a method for a complete characterization of a femtosecond ultraviolet pulse when a fundamental
  near-infrared beam is also available. Our approach relies on generation of second harmonic from the pre-characterized
  fundamental, which serves as a reference against which an unknown pulse is measured using spectral interference (SI).
  The characterization apparatus is a modified second harmonic frequency resolved optical gating setup which
  additionally allows for taking SI spectrum. The presented method is linear in the unknown field, simple and
  sensitive. We checked its accuracy using test pulses generated in a thick nonlinear crystal,
  demonstrating the ability to measure the phase in a broad spectral range, down to 0.1\% peak spectral intensity
  as well as retrieving $\pi$ leaps in the spectral phase.
\end{abstract}


\section{Introduction} \label{sec:intro}
Techniques allowing for a complete characterization of the femtosecond pulse instantaneous intensity and frequency were first
introduced in the nineties. The FROG (Frequency Resolved Optical Gating) \cite{FROGbook}, SPIDER (Spectral
Interferometry for Direct E-field Reconstruction) \cite{IaconisQE99} and sonographic techniques \cite{ReidOL02} are the
most popular. These techniques are usually based on sum frequency generation of an unknown near infrared (NIR) pulse with a modulated copy
of itself. The resultant UV radiation is registered as a function of modulation. Finally, the pulse envelope and phase are
retrieved from acquired data. It is crucial for those methods to filter out the fundamental components from the
frequency sum signal. Fortunately it can be easily accomplished by inserting a color glass filter. Moreover the UV
signal is measurable with inexpensive silicon-based photodetectors.

Frequently one faces the need to characterize the second harmonic (SH) of the pulse of Ti:Sapphire lasers. This is the
case in the ultrafast spectroscopy \cite{tellelaserchem,femtosecondlaserspectroscopy}, micromachining
\cite{liuIEEJQE33} or downconversion-based photon pair sources \cite{pittmanPRA66,URenQIC03,Dragan2004,Keller97}.
Directly generalized FROG and SPIDER techniques suitable for characterization of the UV pulses are based on upconverting
or downconverting the unknown second-harmonic pulse using the fundamental beam \cite{LindenOL99,LonderoJMO03}. The
resultant radiation, which carries information about the unknown pulse, is difficult to detect. In the case of using
upconversion it is centered around 266~nm where the silicon detectors are inefficient. On the other hand using
downconversion produces weak pulses at the fundamental wavelength that are easily overwhelmed by the stray background.
Moreover the nonlinear up- or downconversion process requires substantial intensity of the measured beam. The latter
requirement has recently been diminished by application of an optical parametric amplification based conversion process
\cite{ZhangOE03,TADPOLEOL96}, however these new approaches are complicated in use.
	\begin{figure}
		\centering
		\includegraphics[width=0.50\textwidth]{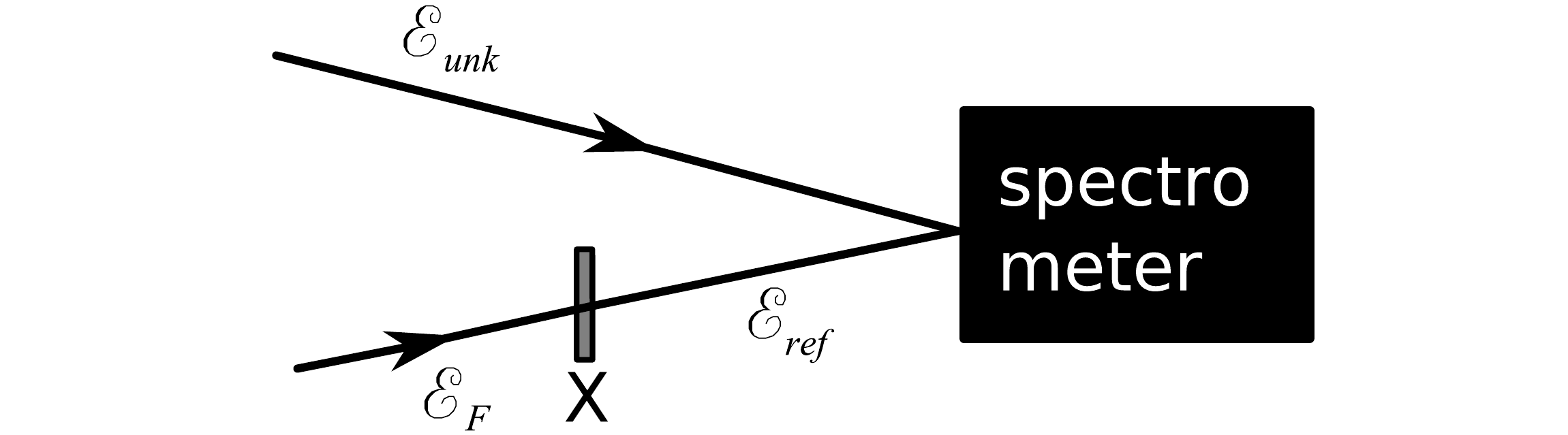}
		\caption{Pictorial diagram of the measurement scheme. $\mathcal E_{\text{unk}}$: unknown pulse,
        $\mathcal E_{\text{F}}$: fundamental pulse, $\mathcal E_{\text{ref}}$: reference second harmonic,
        X: thin nonlinear crystal.}
		\label{fig:schemat}
	\end{figure}

In this paper we present a method for complete characterization of SH pulses when the fundamental beam is also
available. This is a typical experimental situation, which can be exploited to avoid nonlinear conversion of a SH beam.
It is assumed, that the fundamental pulses are characterized prior to the measurement, for example using the
well-established second harmonic generation frequency optical gating (SHG FROG) technique \cite{FROGbook}. The scheme of our method is presented in Fig. \ref{fig:schemat}. A
portion of the fundamental beam is converted into reference second harmonic. Thanks to application of a very thin
crystal in this step the envelope of the reference pulse can be precisely calculated. Then the reference and the
unknown pulse are brought to interfere on a slit of a spectrometer and spectral fringes are registered. The phase of
the unknown pulse is retrieved from this signal using the Fourier filtering technique \cite{takeda_josa72}. We
demonstrate this method for a pulse generated in 1~mm thick beta barium borate (BBO) crystal oriented for type I phase matching. We were able to
reconstruct the phase in spectral range where the pulse intensity is above 0.1\% maximum, including the $\pi$ phase
leaps.

\section{Method}
The first step in our measurement method is registering FROG trace for the fundamental beam \cite{FROGbook}. Next
we run the standard retrieval procedure using FROG software {Femtosoft Technologies) and obtain a complete
information on the electric field of the fundamental pulses. In particular we learn about their spectral envelope
$\mathcal{E}_{\text{F}}(\omega)$:
\begin{equation}
\mathcal{E}_{\text{F}}(\omega) = \sqrt{I_{\text{F}}(\omega)}e^{i\phi_{\text{F}}(\omega)},
\end{equation}	
where $I_{\text{F}}(\omega)$ is the spectral intensity and $\phi_{\text{F}}(\omega)$ is the spectral phase. Note that $I_{\text{F}}(\omega)$
can be measured directly with a spectrometer. This gives us a possibility to verify the accuracy of the data retrieved
from FROG. Next we double known fundamental pulses in a BBO crystal. The
crystal is thin as compared to the distance on which NIR and UV pulse diverge in time by their duration due to
crystal dispersion. We take advantage of this fact to calculate precisely the output second harmonic and use it as a reference. It's spectral envelope $\mathcal{E}_{\text{ref}}(\omega)$ is given by the usual convolution formula:
\begin{equation} \label{convo}
\mathcal{E}_{\text{ref}}(\omega) \propto \int d \omega ' \mathcal{E}_\text{F} (\omega')\mathcal{E}_\text{F}(\omega - \omega').
\end{equation}
In particular we learn about the spectral phase of the reference pulses $\phi_{\text{ref}}(\omega)$. Note that above
formula is a reflection of the fact that the second harmonic field in time is the fundamental field squared:
$\mathcal{E}_{\text{ref}}(t) = \mathcal{E}_\text{F}^2(t)$.

We exploit the spectral interference method in order to retrieve  the difference of phase between the unknown second
harmonic pulse $\mathcal{E}_{\text{unk}}(\omega)$ and the reference pulse $\mathcal{E}_{\text{ref}}(\omega)$. This is accomplished by
directing them with a relative delay $\tau$ into the spectrometer slit, where they interfere. Hence we measure the
interference spectrum $I_{SI}(\omega)$ of the form:
\begin{eqnarray}
\label{ISI}
I_{SI}(\omega)
& = & \left| \mathcal{E}_{\text{ref}}\left(\omega\right)e^{-i \omega \tau} + \mathcal{E}_{\text{unk}} \left(\omega \right) \right|^2    \nonumber\\
& = & I_{\text{ref}}\left(\omega\right) + I_{\text{unk}}\left(\omega \right) \nonumber\\
&   & + 2 \sqrt{I_{\text{ref}}\left(\omega \right)I_{\text{unk}}\left(\omega \right)} \nonumber\\
&   & \quad\times\cos \left(\phi_{\text{ref}}(\omega)-\phi_{\text{unk}}(\omega) - \omega\tau \right).								
\end{eqnarray}
An exemplary interference spectrum $I_{\text{SI}}(\omega)$ is shown in Fig.~\ref{fig:prazki70}. The key information about the
unknown phase $\phi_{\text{unk}}(\omega)$ is contained in the $\cos \left(\phi_{\text{ref}}(\omega)-\phi_{\text{unk}}(\omega) - \omega\tau
\right)$ term. First, we retrieve $\phi_{\text{ref}}(\omega)-\phi_{\text{unk}}(\omega)$ using the Fourier filtering technique
\cite{takeda_josa72}. The interference spectrum can be separated into three distinct elements: the sum of spectral
intensities $I_{\text{ref}}\left(\omega\right) + I_{\text{unk}}\left(\omega \right)$ which is slowly varying with $\omega$ and the
fringes described by the last term in (\ref{ISI}), which consist of positive and negative frequency parts varying as
$e^{-i\omega \tau}$ and $e^{-i\omega \tau}$. We calculate the Fourier transform of the interference spectrum, which
splits those parts in the Fourier domain, provided the delay $\tau$ is large enough. A typical result is shown in Fig.~\ref{fig:fft}.
Next, the Fourier transform is multiplied by a supergauss filter function $F(\tilde{t})$:
\begin{equation}
 F(\tilde{t}) = \exp \left[ -\left(\frac{\tilde{t} - \tau}{\Delta t}\right)^8 \right],
 \label{supergauss}
\end{equation}
where $\tilde{t}$ denotes time in the Fourier domain and $\Delta t$ is the filter width. $F(\tilde{t})$ is nonzero near the
$\tilde{t}=\tau$ point and zero elsewhere. Thus the product of the filter function $F(\tilde{t})$ and the interference
spectrum in the Fourier domain $\tilde I_{SI}(\tilde t)$ contains only the term originating from the part of
$I_{\text{SI}}(\omega)$ which varies as $e^{+i\omega \tau}$. It is transformed back into the frequency domain, yielding:
\begin{eqnarray}
I_{\text{SI}}^{(+)}(\omega) &=& \sqrt{I_{\text{ref}}\left(\omega \right)I_{\text{unk}}\left(\omega \right)} \nonumber \\
&& \times \exp \left[i\left(\phi_{\text{ref}}(\omega)-\phi_{\text{unk}}(\omega) + \omega\tau \right)\right]
\end{eqnarray}
The argument of $I_{\text{SI}}^{(+)}(\omega)$ is the phase difference $\phi_{\text{unk}}(\omega) - \phi_{\text{ref}}(\omega)$ and the linear
term $\omega\tau$. Hence the unknown phase is given by the following formula:			
\begin{equation}
\phi_{\text{unk}} = \arg\left( I_{\text{SI}}^{(+)}(\omega) \right) + \phi_{\text{ref}} + \omega \tau.
\end{equation}	
Finally the spectral intensity of the unknown pulse $I_{\text{unk}}(\omega)$ is measured with a spectrometer and the spectral
envelope can be reconstructed, which concludes the characterization:
\begin{equation}
\mathcal{E}_{\text{unk}}(\omega) = \sqrt{I_{\text{unk}}(\omega)}e^{i\phi_{\text{unk}}(\omega)}.
\end{equation}

Note, that in the experiment nearly identical pattern of the fringes is obtained for a given magnitude of the
relative delay $\tau$ between the pulses, regardless of its sign. On the contrary, the above formulas remain valid only
when a correct sign of $\tau$ is used: from the convention used in Eq.~\eqref{ISI} it follows that if the reference
beam path is delayed then positive $\tau$ is to be inserted. Also, note that a small variation in the assumed value of
$\tau$, difficult to avoid in any real experiment, results only in an irrelevant temporal shift of the reconstructed unknown pulse
shape.

\begin{figure}
    \centering\includegraphics[width=0.50\textwidth]{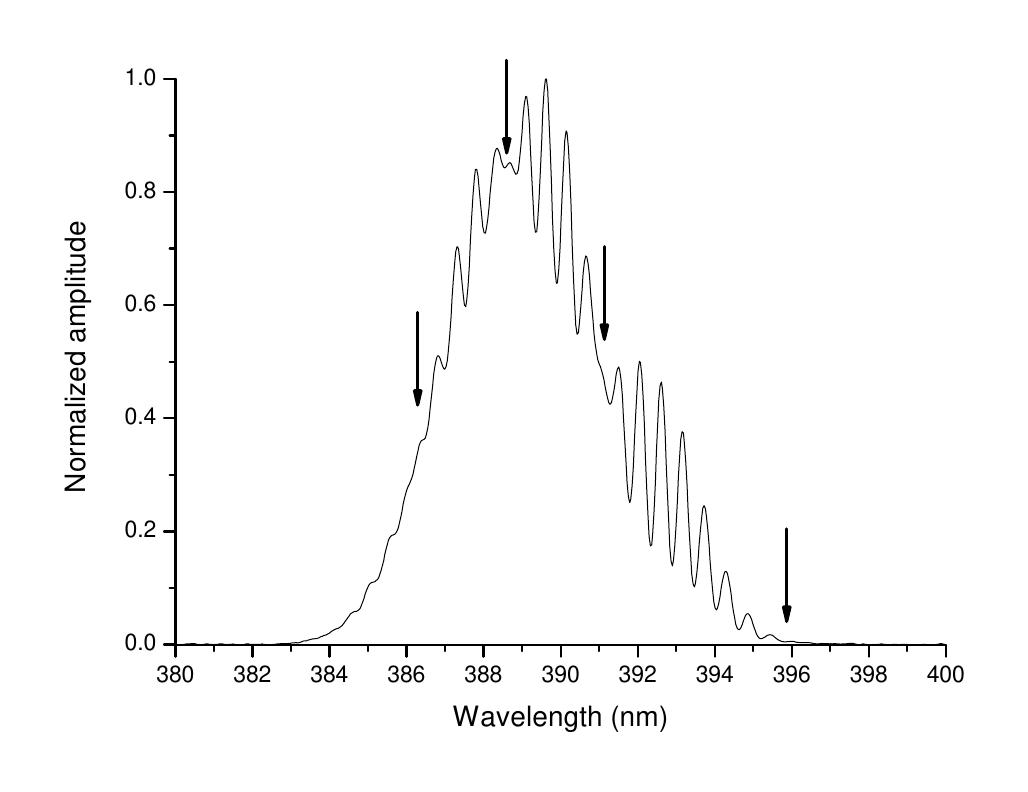}
    \caption{Interference fringes measured by the spectrometer used in the method.
    					For a detailed description of the setup see Sec. \ref{sec:setup}. Arrows mark points of $\pi$ spectral phase leaps.}
    \label{fig:prazki70}
\end{figure}
\begin{figure}
	\centering\includegraphics[width=0.50\textwidth]{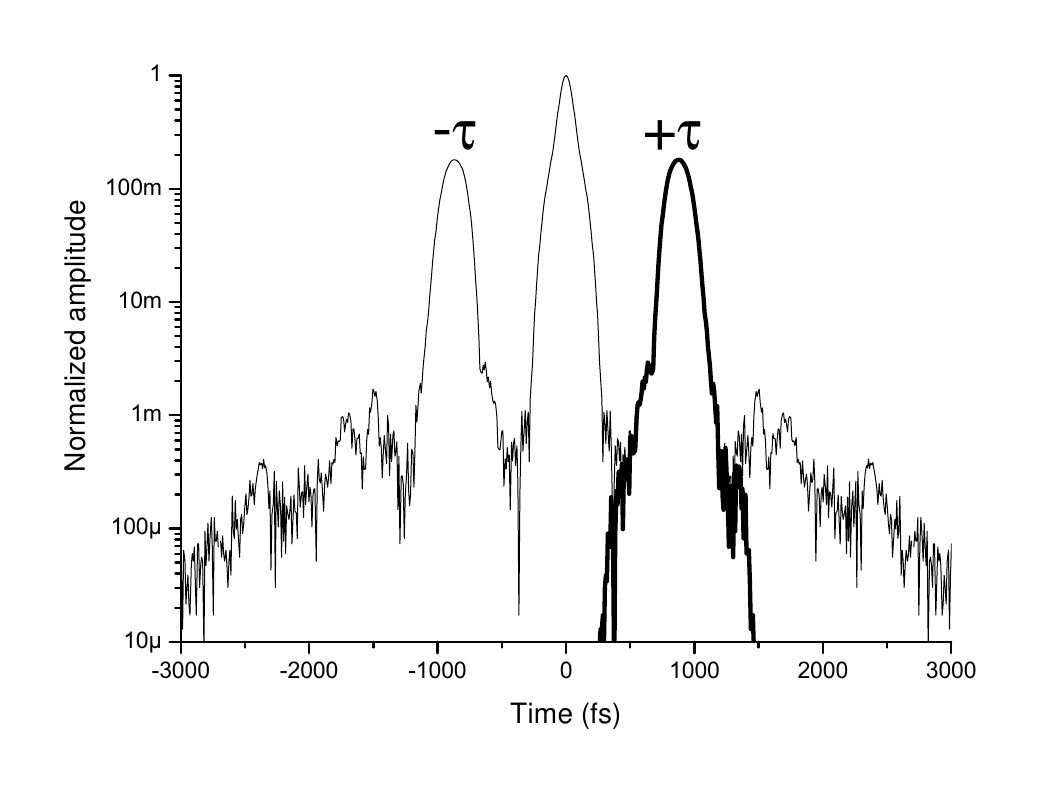}
	\caption{The Fourier transform of the interference fringes pattern (thin line) and the filtered signal (thick) obtained using
						the supergauss filter form Eq. \ref{supergauss} with filter width $\Delta t=500~\text{fs}$
						and temporal delay $\tau = 1000~\text{fs}$. Note the logarithmic scale on vertical axis.}
	\label{fig:fft}
\end{figure}
	
\section{Measurement setup} \label{sec:setup}
Our measuring apparatus is depicted in Fig. \ref{fig:wiazki3}. A 780~nm centered beam from a mode-locked Ti:sapphire laser is
split on a beamsplitter BS1 into two paths. One part goes directly to the FROG-SI measurement apparatus, while the
other is sent to the test pulse preparation arm.

In the first step we measure the SHG FROG trace of the fundamental beam. This measurement requires two replicas of the
fundamental beam, which are produced by the beamsplitter BS2. One replica reflected off BS2 undergoes an adjustable delay by
bouncing off the cornercube mirror CCM, while the second one, transmitted through BS2, is directed with help of 
the flipping mirror FM.
Both thus produced pulses are focused on a 0.05~mm thick BBO type I crystal X where the second harmonic
generation occurs and three beams emerge. Two of them: upper and lower are blocked by a diaphragm D. The middle beam
is filtered out of a scattered fundamental beam by a blue filter BF and focused with a lens L2 on the spectrometer slit.
We retrieved the fundamental pulse with a FROG error \cite{FROGbook} of 0.0012 and we checked that the retrieved spectrum is consistent
with the fundamental beam spectrum measured directly by the spectrometer. 	


The second step is to measure the interference of the reference pulse and the unknown UV pulse generated by an experiment.
For this purpose we switch the position of the flipping mirror to direct the generated UV pulse to the spectrometer.
This pulse is assumed to pass unaffected through the crystal X, while the NIR beam reflected off the cornercube mirror CCM
generates second harmonic of the form given by Eq. \ref{convo}. These two UV pulses are let through by removing the diaphragm D
and made interfere on the slit of the spectrometer.

Note that the crystal suitable for FROG must be 
thin enough to guarantee the validity of Eq. \ref{convo}. 
Therefore we can use the same crystal for reference pulse generation as for SHG FROG measurement.
\begin{figure}
	\centering\includegraphics[width=0.50\textwidth]{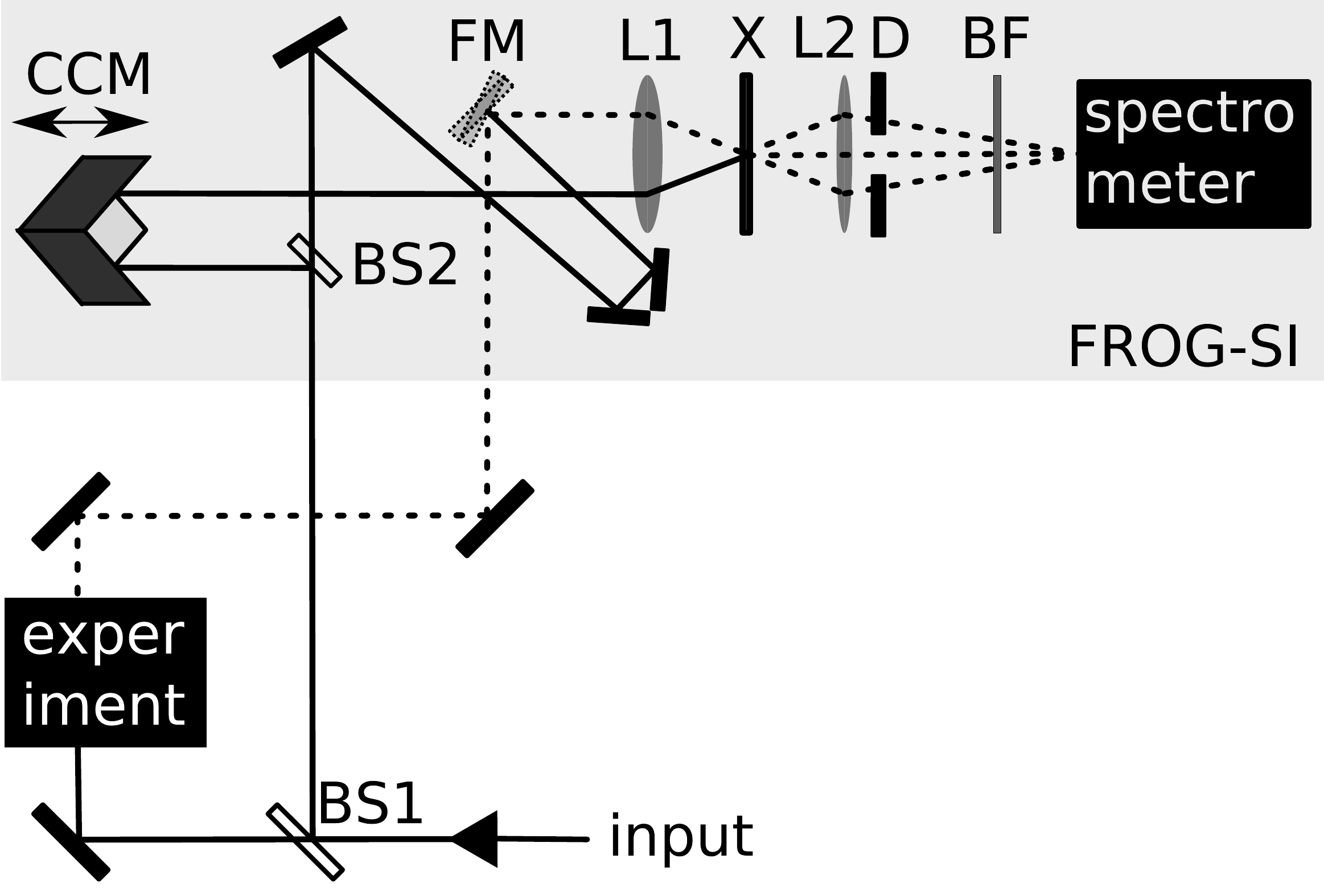}
	\caption{Experimental setup. BS1, BS2: beamsplitters; FM, CCM: mirrors; L1, L2: lenses;
    X: SHG crystal; D: diaphragm; BF: blue filter. }
	\label{fig:wiazki3}
\end{figure}

\section{Results}
We verified the accuracy of our method using test pulses generated in a 1~mm thick BBO crystal and filtered out of remaining NIR
light and attenuated with neutral density filters. The crystal and filters are represented as an ``experiment'' in Fig. \ref{fig:wiazki3}. The test pulse is directed to the FROG-SI apparatus.
When producing the test pulses we did not focus the fundamental beam in the
crystal to achieve a well defined phase matching angle. This simplified the calculation of the test pulse envelope and
made the spectral phase features more pronounced.
  	
In Fig.~\ref{fig:faza} we plot a comparison of a theoretically calculated pulse and one retrieved experimentally. The
results displayed in Fig. \ref{fig:faza}(a) were obtained for thick crystal oriented for phase matching at the wavelength
778~nm, which is equal to the
central wavelength of the incident fundamental beam. Fig.~\ref{fig:faza}(b) shows the result of a similar measurement with the crystal
tilted by 63~mrad, which results in phase matching at 785~nm. Note the fidelity of $\pi$ phase leaps at wavelengths where the
spectral intensity meets zero. The leaps can be noticed in raw interference fringes Fig.~\ref{fig:prazki70}. The
inversion procedure naturally fails at wavelengths where the unknown pulse intensity is below the noise level and the fringes
disappear. From Fig.~\ref{fig:faza} we infer that the threshold spectral intensity above which phase reconstruction is possible is about 0.1\% of the maximum. Note that 
our method is capable of retrieving phase in case when the spectrum of the unknown pulse is nonzero
in two or more separated regions.

\begin{figure}
	\includegraphics[width=0.6\textwidth]{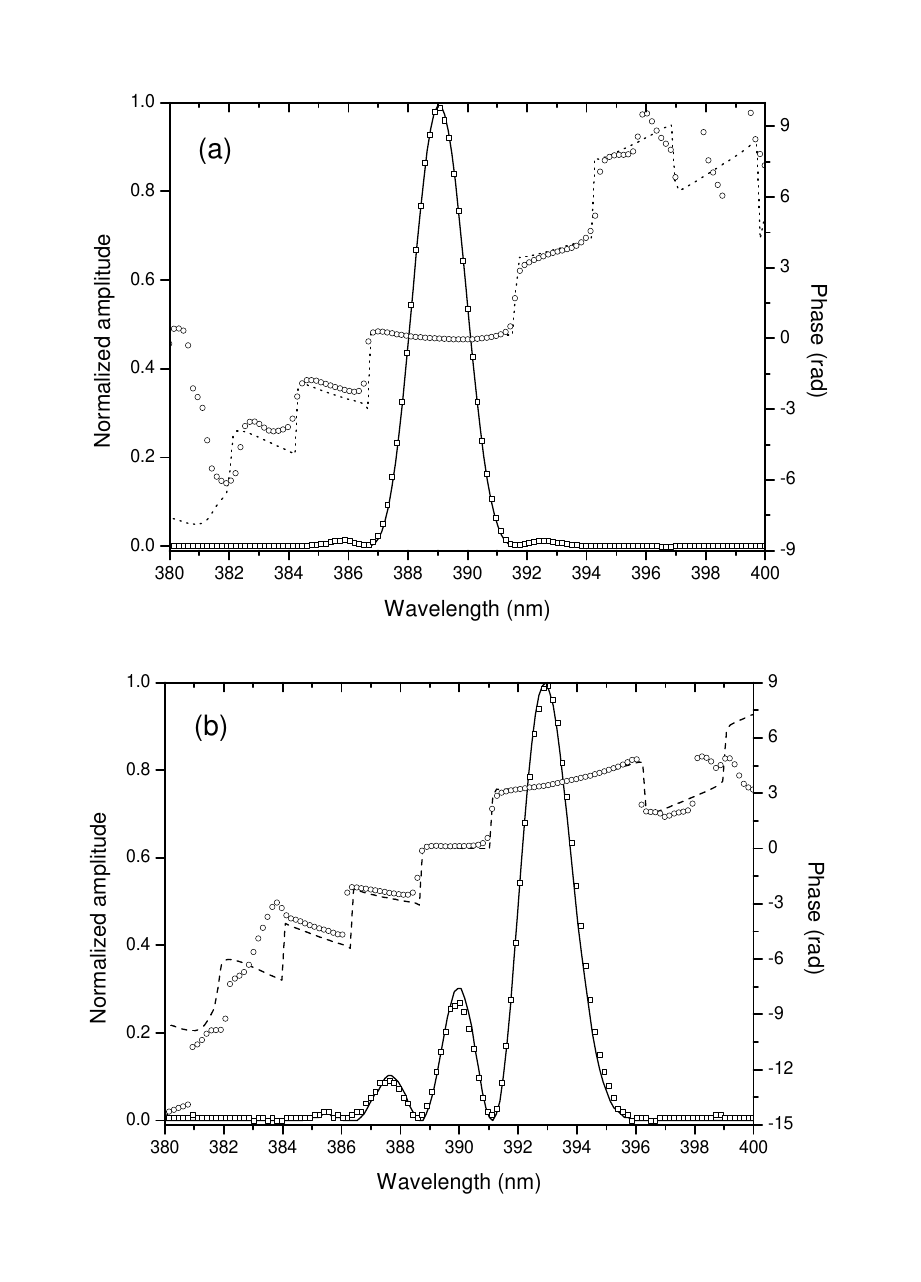}
	\caption{Spectral amplitude (solid line-simulated, rectangles - retrieved experimentally) and phase (dashed line - simulated, circles - retrieved)
	obtained for thick crystal oriented for phase matching at 778~nm (a) and with the crystal tilted by 63~mrad which results in phase matching at 785nm (b).}
    \label{fig:faza}
\end{figure}

We calculated the pulse intensity and phase at the output face of the crystal producing test pulses using the formula for
second harmonic $\mathcal{E}_{\text{SH}}(\omega)$ generated in a
thick nonlinear medium in case when pulse distortion can be neglected
\cite{WasilewskiAPB04}:
\begin{eqnarray}\label{Eq:ESH=sincint}
\mathcal{E}_{\text{SH}}(\omega)& \propto & \mathrm{sinc} \left(\frac{ \Delta\beta L}{2}(\omega-2\omega_{pm}) \right)   \nonumber\\
&& \times \int d\omega' A_1(\omega') A_2(\omega - \omega')
\end{eqnarray}
where $L$ is the crystal thickness, $\omega_{pm}$ is the frequency of perfect phase matching  and $\Delta\beta$ is the
difference of the reciprocals of group velocities of fundamental and second harmonic waves.

The necessary presence of neutral and color glass filters in the test pulse path influences it by contributing an
additional phase $\Delta\phi_{\text{filter}}(\omega)$. We describe it by Taylor expansion around the pulse central frequency
$\omega_0$:
\begin{equation}
\Delta\phi_{\text{filter}}(\omega) = \beta_{0f} + \beta_{1f} (\omega - \omega_0) + \frac{\beta_{2f}}{2} (\omega - \omega_0)^2 + ...
\end{equation}		
The constant and linear phase terms $\beta_{0f}$ and $\beta_{1f}$ correspond to pulse retardation. The only relevant
and appreciable contribution is the quadratic dispersion $\beta_{2f}/2 (\omega - \omega_0)^2$.

For measuring $\beta_{2f}$ we replaced a 1~mm BBO crystal with one that was 0.1~mm thick, which generated more broadband test
pulses. Next, we acquired a reference spectral interferogram. Then we inserted additional filters, identical to those to
be characterized, in the test path and registered the interference again. The difference of the spectral phases
retrieved in those two cases equals $\Delta\phi_{filter}(\omega)$. By fitting the latter by a square polynomial we
computed $\beta_{2f}$. Using these data we could substract the influence of the filters from the measured test pulse phase
obtaining the result shown in Fig. \ref{fig:faza}.


The agreement of measurement results with theoretical models certifies the validity of the presented method.
														
\section{Summary}
We have demonstrated a novel method for a complete characterization of the near-UV second harmonic of ultrashort
pulses.	It is applicable in situations when an intense NIR beam is available. Additionally, the unknown pulse
spectrum must lie within the spectral range of second harmonic generated in a very thin crystal.

The main advantage of the presented method is a simple setup. It is a hybrid of standard SHG FROG with a spectral
interferometer and there is no need to employ additional spectrometer or crystals. The pulse shape is inferred
from an UV spectrum which is easily registered. The algorithm used for phase retrieval is simple and non-iterative.	
In addition, our method is linear in the unknown pulse field.

The interference acquisition is single shot, without a need to scan a range of time delays $\tau$. Additional averaging
over various $\tau$ is possible with compensation of a linear phase.

Finally, our method allowed us to characterize rather small phase contributed to the UV pulses by glass filters.
Measurement of such a small phase would be very difficult to achieve using FROG.

\section*{Acknowledgements}
We acknowledge insightful discussions with C. Radzewicz, \L. Kornaszewski and P. Wasylczyk. This work has
been supported by the Polish budget funds for scientific research projects in years 2006-2008, the European Commission
under the Integrated Project Qubit Applications (QAP) funded by the IST directorate as Contract Number 015848.
It has been carried out in the National Laboratory for Atomic, Molecular, and
Optical Physics in Toru\'{n}, Poland.

\end{document}